\documentclass[a4paper,11pt]{article}
\usepackage{pos}
\usepackage{amsmath}
\usepackage{dsfont}
\usepackage{bbm}
\usepackage{cleveref}
\usepackage{subcaption}
\usepackage[utf8]{inputenc}

\let\OLDthebibliography\thebibliography
\renewcommand\thebibliography[1]{
  \OLDthebibliography{#1}
  \setlength{\parskip}{0pt}
  \setlength{\itemsep}{0pt plus 0.3ex}
}

\DeclareMathOperator{\tr}{tr}

\newcommand{\fm}{\text{fm}}
\newcommand{\dif}{\mathrm{d}}

\newcommand{\sys}{\mathrm{sys}}
\newcommand{\flow}{\mathrm{fl}}

\newcommand{\chihat}{\hat\chi}
\newcommand{\rhat}{\hat r}
\newcommand{\ahat}{\hat a}
\newcommand{\tflow}{t_{\flow}}

\title{The influence of gauge field smearing on discretisation effects}
\ShortTitle{The influence of gauge field smearing on discretisation effects}

\author*[a]{Andreas Risch}
\author[a]{Stefan Schaefer}
\author[a,b]{Rainer Sommer}

\affiliation[a]{John von Neumann-Institut f{\"u}r Computing NIC, Deutsches Elektronen-Synchrotron DESY,\\
Platanenallee 6, 15738 Zeuthen, Germany}
\affiliation[b]{Institut f{\"u}r Physik, Humboldt-Universit{\"a}t zu Berlin\\
Newtonstr. 15, 12489 Berlin, Germany}

\emailAdd{andreas.risch@desy.de}
\emailAdd{stefan.schaefer@desy.de}
\emailAdd{rainer.sommer@desy.de}

\abstract{When designing lattice actions, gauge field smearing is frequently used to define the lattice Dirac operator. Since the smearing procedure removes effects of ultraviolet fluctuations, the fermions effectively see a larger lattice spacing than the gauge fields. Creutz ratios, formed from ratios of rectangular Wilson loops, based on smeared gauge fields are adequate observables to investigate the effect of smearing since they do not need renormalisation and provide a measure of the physical forces felt by the fermions. We study their behaviour at various smearing radii (fixed in lattice units) and in particular how the smearing influences the scaling towards the continuum limit. Since we employ the Wilson gradient flow as smearing, the same Creutz ratios have another, well defined continuum limit, when the flow time is fixed in physical units. That continuum limit is reached with smaller corrections at finite $a$.}

\FullConference{%
The 39th International Symposium on Lattice Field Theory,\\
8th-13th August, 2022,\\
Rheinische Friedrich-Wilhelms-Universit{\"a}t Bonn, Bonn, Germany
}

\begin{document}
\maketitle

\section{Introduction}

Four-dimensional smearing is a common method used in lattice gauge theories to smooth a gauge field. During the years, several smearing algorithms have been developed, e.g. HYP~\cite{Hasenfratz:2001hp}, Stout~\cite{Morningstar:2003gk}, HEX~\cite{Capitani:2006ni} and gradient flow~\cite{Narayanan:2006rf,Luscher:2010iy} smearing. Considering a smearing transformation $\mathcal{S}:U\mapsto \mathcal{S}[U]$ there are two main types of application: Observable smearing
\begin{gather}
\langle O_{\mathcal{S}}[U] \rangle = \langle O[\mathcal{S}[U]] \rangle,
\end{gather}
where a new observable is defined by evaluating a given observable on the smeared gauge field, and smearing in the lattice fermion action
\begin{gather}
S[U] = S_{\mathrm{g}}[U] + \overline{\psi}\,D[\mathcal{S}[U]]\,\psi,
\end{gather}
where the Dirac operator is evaluated on the smeared gauge field. Concerning observable smearing the main motivation is noise reduction, whereas smearing in the fermion action improves the algorithmic stability, e.g. with regard to exceptional configurations. In Ref.~\cite{Hasenfratz:2007rf} even at very coarse lattice spacings the Wilson Dirac operator defined with nHYP gauge links could be shown to exhibit a spectrum with a well-defined spectral gap. The same was shown for stout smearing in Ref.~\cite{Durr:2008rw}.

In this work we are going to focus on two questions: How does the smearing strength influence lattice artefacts? What smearing strengths allow for a controlled continuum extrapolation? We will discuss these issues using the example of Creutz ratios, which provide a measure of the physical\footnote{As opposed to the forces present in the molecular dynamics evolution of the HMC.} forces felt by the fermions caused by the gauge field. A somewhat related discussion for thermal correlation functions and the energy-momentum tensor can be found in~\cite{Eller:2018yje}.  Creutz ratios in combination with APE smearing and the Wilson flow were also studied in a determination of the string tension in~\cite{Okawa:2014kgi}.

\section{The gradient flow formalism, gradient flow smearing and physical gradient flow}

We make use of the continuum four-dimensional Yang-Mills action which is defined as $S_{\mathrm{YM}}=-\frac{1}{2g_0^2}\int \dif^4 x \tr( F_{\mu\nu}(x)F_{\mu\nu}(x))$, where $F_{\mu\nu} = \partial_{\mu}A_{\nu}-\partial_{\nu}A_{\mu}+[A_{\mu},A_{\nu}]$ denotes the field strength tensor and $A_{\mu}(x)$ the corresponding gauge field. In the Yang-Mills continuum gradient flow formalism~\cite{Luscher:2010iy} a gauge field $B_{\mu}(x,\tflow)$ is introduced, which is defined on $\mathds{R}^4\times [0,\infty)$ and where $\tflow$ is the so called gradient flow time. At $\tflow=0$ the usual four-dimensional gauge field is an initial condition in the flow time evolution, i.e.
\begin{gather}
B_{\mu}(x,0) = A_{\mu}(x).
\end{gather}
The dynamics in the flow time direction is driven by the gauge-covariant flow equation
\begin{gather}
\frac{\partial}{\partial \tflow}B_{\mu}(x,\tflow) = -\frac{\delta S_{\mathrm{YM}}[B]}{\delta B_{\mu}(x,\tflow)} = D_{\nu}G_{\nu\mu}(x,\tflow),
\end{gather}
where $G_{\mu\nu} = \partial_{\mu}B_{\nu}-\partial_{\nu}B_{\mu}+[B_{\mu},B_{\nu}]$ denotes the generalised field strength tensor on $\mathds{R}^4\times [0,\infty)$ and $D_{\mu} = \partial_{\mu} + [B_{\mu}, \cdot]$ the generalised covariant derivative. The gradient flow possesses a smoothing property in the direction of positive flow time: From a leading-order perturbative expansion one may deduce that the gauge field $B_{\mu}(x,\tflow)$ is a spherically smoothed version of $A_{\mu}(x)$ with mean-square radius $r_{\mathrm{sm}} = \sqrt{8 \tflow}$~\cite{Luscher:2010iy}. In~\cite{Luscher:2011bx} it has been shown perturbatively to all loop orders that {\em any} functional of the flowed fields $B_{\mu}(x,\tflow)$ at strictly positive $\tflow$ is finite, i.e. no additional renormalisation is required, assuming that the four-dimensional theory has been renormalised.

We discretise the Yang-Mills gradient flow by means of the Wilson gradient flow~\cite{Luscher:2010iy} and integrate the flow equation numerically by means of an explicit 3rd-order Runge-Kutta integration scheme with a step size $\frac{\Delta \tflow}{a^{2}}\leq$ which never exceeds $
0.01$.

We consider {\bf two scenarios} in which we apply the gradient flow to the gauge field: For gradient flow {\bf smearing} the gradient flow time is fixed in lattice units when varying the lattice spacing, $\frac{8\tflow}{a^{2}} = \mathrm{const}$, i.e. in physical units it shrinks when the lattice spacing is reduced and therefore vanishes in the continuum limit. Hence, the continuum theory is unaltered. Smearing strengths up to $\frac{8\tflow}{a^{2}}=8$ have been used in practice, e.g. in~\cite{Berkowitz:2017opd}. The second scenario we refer to as a {\bf physical gradient flow}. In this case the gradient flow time is fixed in physical units, i.e. $\tflow / t_0= \mathrm{const}$, where $t_0$ may be any physical length scale of the theory, in particular the scale defined in \cite{Luscher:2010iy}. Consequently, the flow time in lattice units grows with shrinking lattice spacing $\frac{8\tflow}{a^{2}} \propto \frac{8t_0}{a^{2}}$ and hence the continuum theory / observables are altered. One motivation for a physical gradient flow is the possibility to define and access new observables.

\section{Lattice setup}

\begin{table}[h]
\begin{center}
\begin{tabular}{|l|lll|ll|l|}
\hline
ensemble & $\beta$ & $T/a$ & $L/a$ & $a\,[\fm]$ & $L\,[\fm]$ & $t_{0}/a^{2}$ \\
\hline
sft1 & 6.0662 & 80 & 24 & 0.0834(4) & 2.00(1) & 3.990(9) \\
sft2 & 6.2556 & 96 & 32 & 0.0624(4) & 2.00(1) & 7.070(17) \\
sft3 & 6.5619 & 96 & 48 & 0.0411(2) & 1.97(1) & 16.52(6)\\
sft4 & 6.7859 & 192 & 64 & 0.0312(2) & 2.00(1) & 29.60(10)\\
sft5 & 7.1146 & 320 & 96 & 0.0206(2) & 1.98(2) & 67.94(23)\\
\hline
\end{tabular}
\caption{Parameters of the SU(3) gauge ensembles~\cite{Husung:2017qjz} and computed reference flow time $t_{0}/a^{2}$ in lattice units.}
\label{tbl:ensembles}
\end{center}
\end{table}

We make use of $\mathrm{SU}(3)$ Yang Mills theory gauge ensembles~\cite{Husung:2017qjz} based on the Wilson plaquette action. Temporal open boundary conditions~\cite{Luscher:2011kk} are imposed to alleviate topology freezing. The scale is set by means of the force parameter $r_{0}$~\cite{Sommer:1993ce}, where for illustration a  value of $r_{0} = 0.5\,\fm$ is used. An overview of the gauge ensembles is given in \cref{tbl:ensembles}. The lattice spacing varies between $0.08\,\fm$ and $0.02\,\fm$ and the spatial extent is kept constant at $L=2\,\fm$. 

Instead of the force parameter $r_{0}$ we will in the following use the already mentioned reference flow time $t_{0}$~\cite{Luscher:2010iy} to construct dimensionless quantities. To define $t_{0}$ we make use of the action density $E(x,\tflow) = -\frac{1}{2}\sum_{\mu,\nu}\tr\big(G^{\mathrm{clv}}_{\mu\nu}(x,\tflow)\,G_{\mu\nu}^{\mathrm{clv}}(x,\tflow)\big)$, where $G^{\mathrm{clv}}$ denotes the field strength tensor in the clover discretisation~\cite{Sheikholeslami:1985ij}. The reference flow time $t_{0}$ is then implicitly defined by $t^{2}_{0}\,\langle E(x,t_0) \rangle  = 0.3$~\cite{Luscher:2010iy}. Numerical values are listed in \cref{tbl:ensembles}.

In our measurements we implement the two scenarios for scaling the flow time via
\begin{equation}
\label{eq:scenarios}	
\frac{8t_\flow}{a^2} =
\begin{cases}
0,\; 0.25,\; 0.5,\; \ldots,\; 2, \; 2.5, \ldots, \;3.5,\;4,\; 5,\; 6,\; 7, \; 8 & \text{smearing}\\
\frac{8t_0}{a^2} \times0.0146 \times j\,,\;\; j \in\{0,\;1,\;\ldots ,\;4\} & \text{physical flow}
\end{cases}	
\end{equation}

\section{Creutz ratios and gradient flow}

In this work we study the influence of smearing on the continuum extrapolation of Creutz ratios. In order to define the latter we introduce planar rectangular Wilson loops of size $r\times t$,
\begin{gather}
W(r, t) \equiv \Big\langle \tr\Big( P \exp\Big(\oint_{\gamma(r,t)} \dif x_{\mu}A_{\mu}(x)\Big)\Big) \Big\rangle,
\end{gather}
which are obtained from the gauge field by a path-ordered integral along a rectangular closed path $\gamma(r,t)$. The Creutz ratios are then obtained from dual logarithmic derivatives:
\begin{gather}
\chi(r, t) \equiv -\frac{\partial}{\partial t}\frac{\partial}{\partial r} \ln(W(r, t))\label{eq:creutzcont}.
\end{gather}
It can be shown that in the limit of an infinite time extent the static quark anti-quark force can be extracted, i.e. $\chi(r,t) \rightarrow F_{\mathrm{\overline{q}q}}(r)\text{ for }t\rightarrow\infty$~\cite{Okawa:2014kgi}. In order to define Creutz ratios on the lattice~\cite{Creutz:1980wj} we construct planar rectangular Wilson loops from closed paths of gauge links:
\begin{gather}
W^\mathrm{lat}(r,t) \equiv \Big\langle \tr\Big(\prod_{(x,\mu)\in\gamma(r,t)} U_{\mu}(x)\Big) \Big\rangle.
\end{gather}
We discretise \cref{eq:creutzcont} making use of central differences~\cite{Okawa:2014kgi},
\begin{gather}
\chi\Big(t+\frac{a}{2}, r+\frac{a}{2}\Big) \equiv \frac{1}{a^{2}}\ln\Big(\frac{W(t+a, r)\cdot W(t, r+a)}{W(t, r) \cdot W(t+a, r+a)}\Big),
\end{gather}
such that only $O(a^{2})$ lattice artefacts remain. Note that Creutz ratios renormalise trivially, i.e. no extra renormalisation factors are required to obtain a renormalised $\chi(r, t)$. This important property entails that the small flow time limit is smooth, as opposed to e.g. the one of $\langle E(x,t_\flow) \rangle$. For our observables, we can interchange continuum limit and small flow time limit. The two scenarios \cref{eq:scenarios} therefore have a common limit where both $a=0$ and $t_\flow=0$.

In the following discussion, we will only focus on diagonal Creutz ratios $\chi(r, t)$ with $r=t$, which we abbreviate as $\chi(r)\equiv\chi(r, r)$. We compute the latter in lattice units $(\chi\cdot a^{2})(\frac{r}{a})$ for various half integer distances $\frac{r}{a}=1.5,2.5,\ldots$ on gauge configurations on which the gradient flow was applied. We use $t_0$ to define dimensionless Creutz ratios, i.e. we analyse $\chihat \equiv \chi\cdot 8 t_{0}$ as a function of $\rhat \equiv\frac{r}{\sqrt{8t_{0}}}$. The computation is based on the  openQCD~\cite{Luscher:openQCD} package and utilizes B. Leder's program for measuring Wilson loops~\cite{Leder:wloop,Donnellan:2010mx}. For the data analysis the python3 package pyobs~\cite{Bruno:pyobs} is used, which implements the $\Gamma$-method~\cite{Wolff:2003sm} for Monte Carlo error estimation.

\begin{figure}[ht]
\centering
\includegraphics[width=0.495\textwidth]{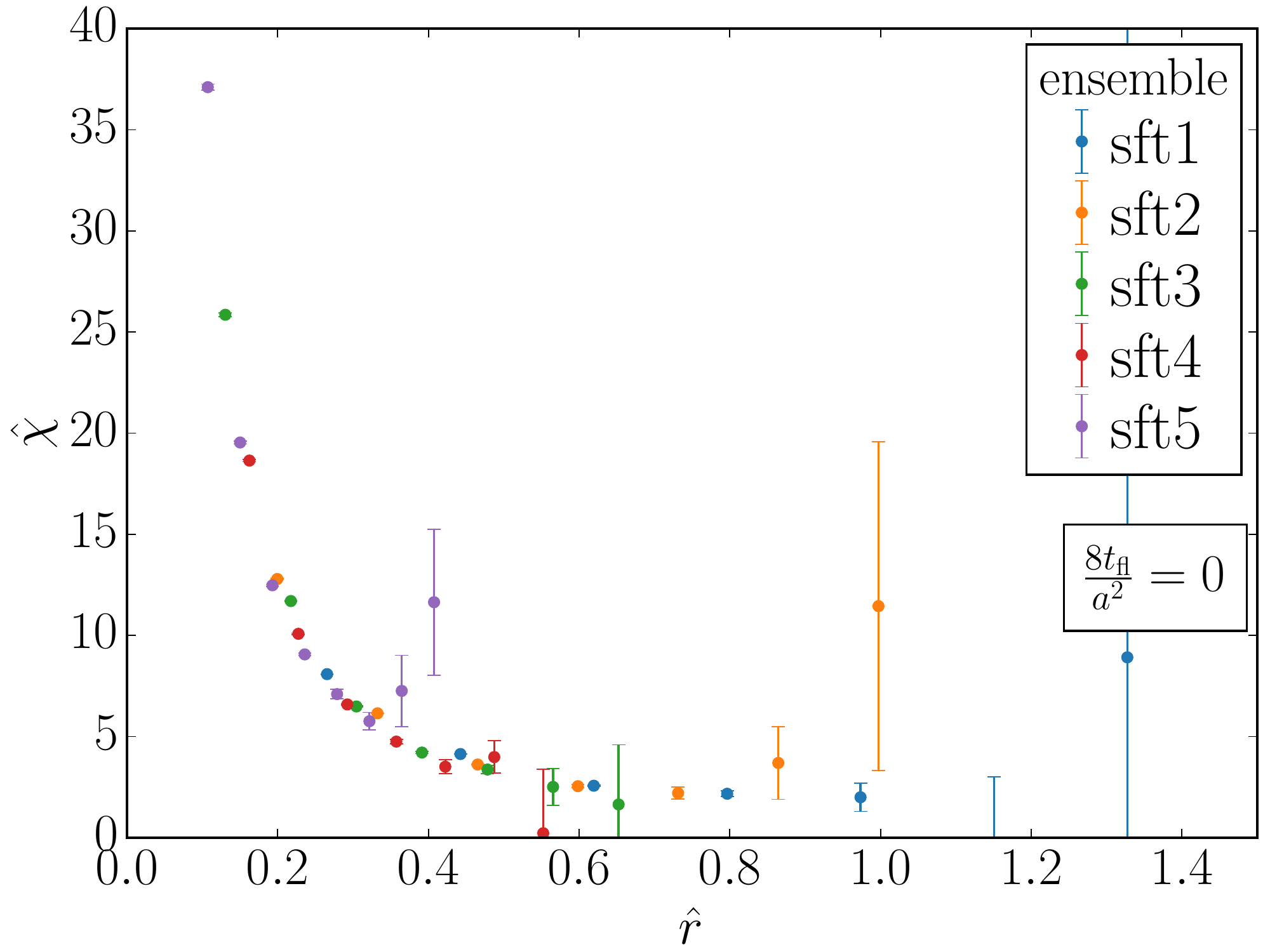}
\includegraphics[width=0.495\textwidth]{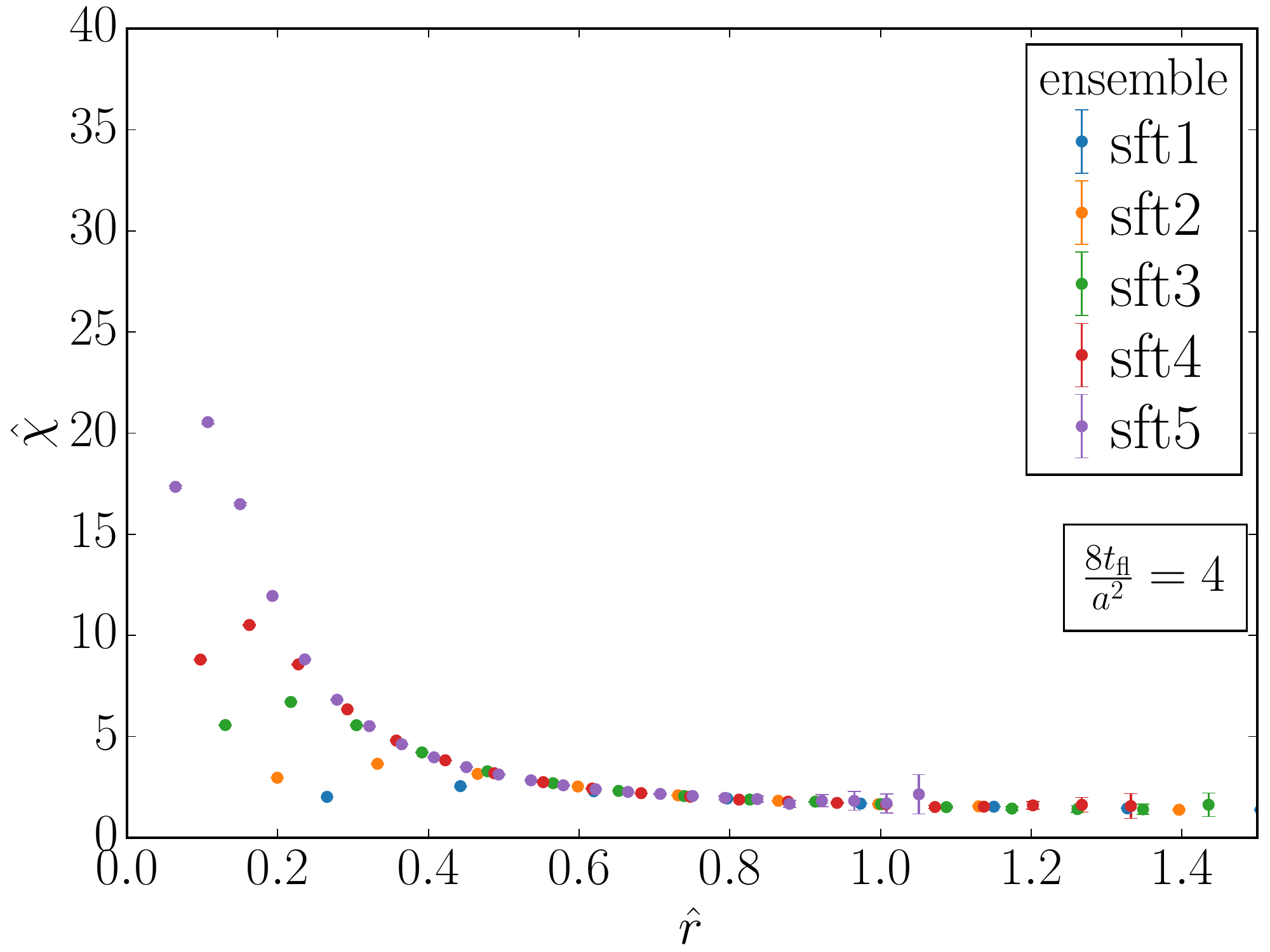}
\caption{Dimensionless Creutz ratios $\chihat\equiv\chi\cdot 8t_{0}$ as functions of the distance $\rhat\equiv\frac{r}{\sqrt{8t_{0}}}$ on different ensembles and for different gradient flow times $\frac{8 \tflow}{a^{2}}$.\label{fig:smearing}}
\end{figure}
In \cref{fig:smearing} the dimensionless diagonal Creutz ratio $\chihat$ is displayed as a function of the distance $\rhat$ evaluated on the five gauge ensembles. For larger distances the statistical signal deteriorates. Comparing both figures, which only differ in the size of the applied gradient flow time $\frac{8 \tflow}{a^{2}}$, we confirm that the gradient flow reduces statistical errors drastically ~\cite{Okawa:2014kgi}. The $\sim \frac{1}{r^2}$ short distance behaviour is smoothed at distances $r \lessapprox \sqrt{8t_\flow}$. Already from this qualitative picture we see that in the smearing scenario the path to continuum and hence lattice artefacts are altered. This effect becomes smaller at larger distances which is coherent with the finiteness of the smearing radius.

\section{Interpolation of Creutz ratios}

We want to extrapolate $\chihat(\rhat)$ to the continuum at a fixed value of $\rhat$. This requires the knowledge of $(\chi\cdot a^{2})\big(\frac{r}{a}\big)$ at arbitrary values of $\frac{r}{a}$, i.e. also at locations $\frac{r}{a}\neq 1.5,2.5,\ldots$, where the values are not determined by the measurements. We therefore interpolate $\chi\cdot a^{2}$ as a function of $\frac{r}{a}$. A first class of considered interpolation models is constructed from linear combinations of $m$ monomials $\big(\frac{r}{a}\big)^{n_{j}}$, where $n_{j}\in\mathbb{Z}$, with coefficients $c_{n_{j}}\in\mathbb{R}$: The function $\mathrm{Pol}_{n_{1},\ldots,n_{m}}\big(\frac{r}{a}\big) = \sum_{j=1}^{m}c_{n_{j}}\big(\frac{r}{a}\big)^{n_{j}}$ is then used to locally interpolate between $m$ adjacent nodes. For $m>2$ several locally defined interpolation functions possess overlapping regions. In order to obtain a smoothing effect for the global interpolation and to resolve the interpolation ambiguity we average over all contributing interpolations in an overlapping region between two adjacent nodes. In particular, we make use of the interpolation models $\mathrm{Pol}_{2,1,0}$ and $\mathrm{Pol}_{0,-2,-4}$. In addition, we use a cubic natural spline $\mathrm{CSpl}$, which is defined as a piecewise cubic polynomial with continuous second derivative and vanishing second derivative at the boundaries, as well as the combination $\mathrm{CSpl}\cdot\mathrm{Pol}_{-2}$. 

\begin{figure}[ht]
\centering
\includegraphics[width=0.5\textwidth]{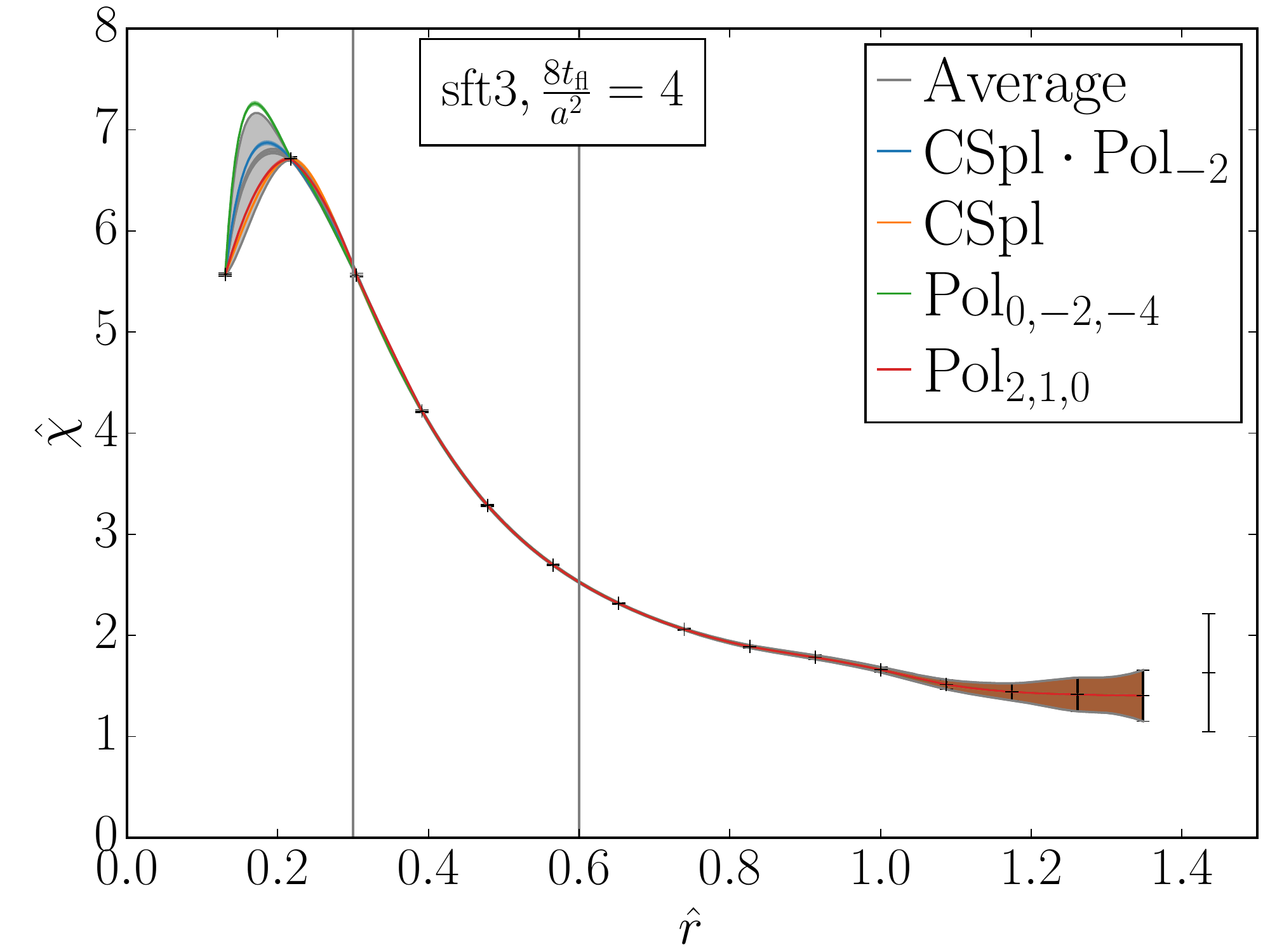}
\caption{Dimensionless Creutz ratio $\chihat\equiv\chi\cdot 8t_{0}$ as a function of the distance $\rhat\equiv\frac{r}{\sqrt{8t_{0}}}$ on the ensemble sft3 with a gradient flow time $\frac{8 \tflow}{a^{2}}=4$ for various interpolation models.}
\label{fig:interpolation}
\end{figure}
In order to account for the interpolation model ambiguities we average over all $n=4$ considered interpolation models $(\chi\cdot a^{2})_{\mathrm{av}}\big(\frac{r}{a}\big) \equiv \frac{1}{n}\sum_{i=1}^{n}(\chi\cdot a^{2})_{i}\big(\frac{r}{a}\big)$ and add a systematic error, which is given by half the spread of all interpolation models: $\Delta_{\sys}(\chi\cdot a^{2})_{\mathrm{av}}\big(\frac{r}{a}\big) \equiv \frac{1}{2}\big(\max_{i=1,\ldots,n}\big\{(\chi\cdot a^{2})_{i}\big(\frac{r}{a}\big)\big\} - \min_{i=1,\ldots,n}\big\{(\chi\cdot a^{2})_{i}\big(\frac{r}{a}\big)\big\}\big)$. In \cref{fig:interpolation} several interpolation models are shown for a given value of the smearing strength on one ensemble. The interpolations begin to differ at short distances in the region where the Creutz ratios vary rapidly between nodes. In the following we are going to focus on the region $0.3 \leq\rhat\leq 0.6$, in which lattice artefacts are not uncontrollably large and the statistical signal is sufficiently good.

\section{Continuum extrapolation}

At fixed distance $\rhat$ we perform a global continuum extrapolation making use of the Symanzik  expansion\footnote{Our investigation has an intermediate precision. At this level we neglect logarithmic effects both in the lattice spacing \cite{Husung:2021mfl} and in the flow time \cite{Luscher:2010iy}.} and the small flow time expansion of $\chihat$. For this purpose we define the dimensionless lattice spacing parameter $\ahat\equiv\frac{a}{\sqrt{8t_{0}}}$ and the flow time parameter $\varepsilon=\frac{\tflow}{t_{0}}$. The double expansion is then given by
\begin{align}
\chihat &= \sum_{i=0}^{n}c_{i} \ahat^{i} + O(\ahat^{n+1}), & c_{i} &= \sum_{i=0}^{m}c_{ij}\varepsilon^{j} + O(y^{m+1}).
\end{align}
To define the fit ansatz for the extrapolation we truncate the latter expansion, obtaining
\begin{align}
\chihat_{\mathrm{tr}}(\ahat,y) &= c_{00} + c_{20} \ahat^{2} + c_{40} \ahat^{4} + c_{01}\varepsilon + c_{21}\ahat^{2}\varepsilon + c_{02}\varepsilon^{2}. \label{eq:fitflow}
\end{align}
In order to demonstrate that this fit ansatz also describes the $a$-dependence for fixed {\bf smearing}, we observe that the smearing strength is parametrised by $\frac{\varepsilon}{\ahat^{2}}=\frac{8\tflow}{a^{2}}$ and
\begin{align}
\chihat_\mathrm{tr} &= d_{0} + d_2  \ahat^{2} + d_{4}\ahat^{4} \label{eq:fitsmearing}
\end{align}
with coefficients
\begin{align}
d_{0}=c_{00}, \quad d_{2}=c_{20} \Big(1+\frac{c_{01}}{c_{20}}\frac{8\tflow}{a^{2}}\Big),\quad d_{4}=c_{40}\Big(1 + \frac{c_{21}}{c_{40}}\frac{8\tflow}{a^{2}} + \frac{c_{02}}{c_{40}}\frac{64\tflow^2}{a^{4}}\Big). 
\end{align}

\begin{figure}[ht]
\centering
\includegraphics[width=0.75\textwidth]{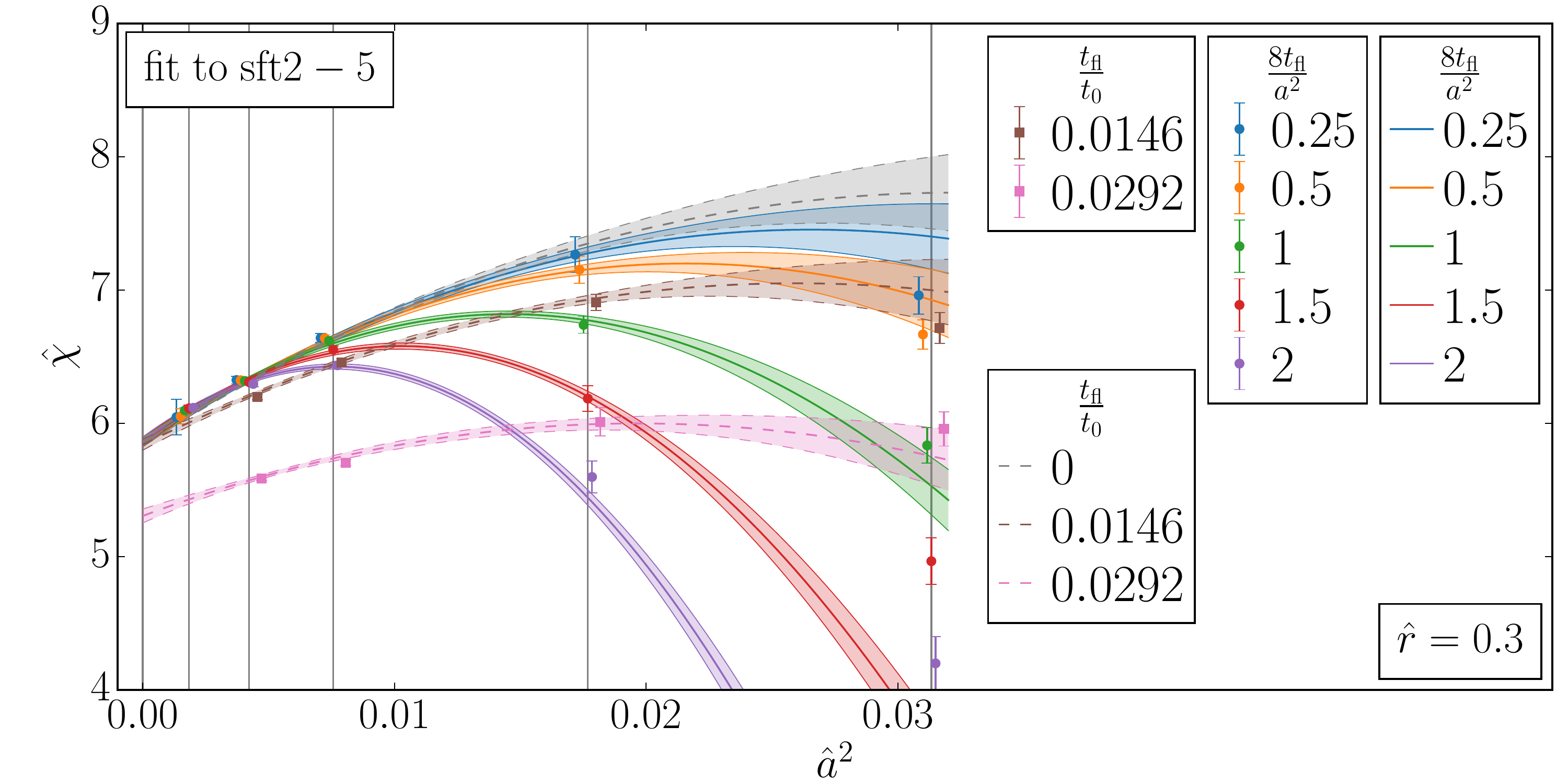} 
\\
\includegraphics[width=0.75\textwidth]{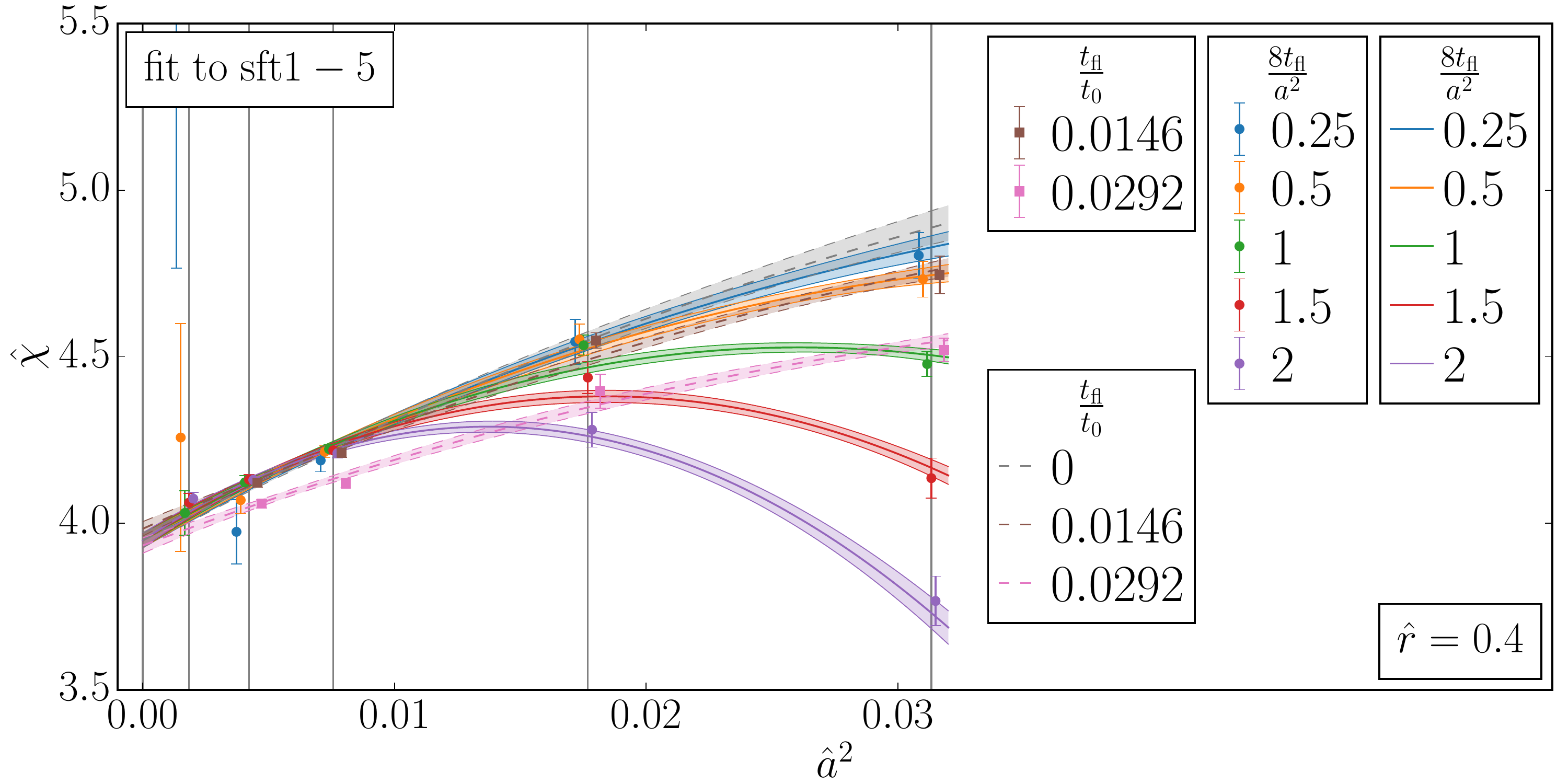}
\caption{Dimensionless Creutz ratio $\chihat\equiv\chi\cdot 8t_{0}$ as a function of the lattice spacing $\ahat\equiv\frac{a}{\sqrt{8t_{0}}}$ at a distance of $\rhat\equiv\frac{r}{\sqrt{8t_{0}}}=0.3$ (top) and $0.4$ (bottom). Extrapolations for several gradient flow smearing strengths $\frac{8\tflow}{a^{2}}$ (solid) and for several physical gradient flows $\frac{\tflow}{t_{0}}$ (dashed). Solid lines and circles belong to gradient flow smearing, whereas dashed lines and squares represent a physical gradient flow. Data points have been shifted for better visibility.}
\label{fig:continuum}
\end{figure}
In \cref{fig:continuum} the continuum extrapolations for the distances $\rhat=0.3$ and $0.4$ are shown. We consider both smearing and the physical gradient flow, c.f. \cref{eq:scenarios}. An inclusion of data from too large smearing strengths $\frac{8\tflow}{a^{2}}$ and too large physical gradient flow strengths $\frac{t_{0}}{\tflow}$ led to  fits with a bad $p$-value, which is an indication of the breakdown of the Symanzik expansion and the small flow time expansion. This depends however on the $\rhat$: at larger distances larger values of $\frac{8\tflow}{a^{2}}$ and $\frac{t_{0}}{\tflow}$ can be fitted with the low-order expansion. By construction of the fit model the continuum limit is independent of the smearing strength $\frac{8\tflow}{a^{2}}$ as in the limit $\ahat\rightarrow 0$ only the fit coefficient $d_{0}=c_{00}$ in \cref{eq:fitsmearing} contributes. In the figures all continuum extrapolations depicted by solid lines, which differ in their smearing strengths, therefore share a common continuum limit. 

In contrast, the {\bf physical gradient flow}, constant $\varepsilon=\frac{\tflow}{t_{0}}$, leads to a flow time dependent continuum limit $c_{00} + c_{01}\varepsilon+c_{02}\varepsilon^2$ which turns out to be below the $c_{00}$ attained with the smearing. This dependence is more pronounced at shorter distances. In contrast to the rather steep bending smearing curves, we observe rather flat extrapolations with the physical flow at constant $t_\flow/t_0$ (dashed). Since the raw data is the same in both cases, the flat behaviour for the physical flow and the increasing difference between the continuum limits for shorter distances directly lead to the non-monotonic and relatively strong dependence of the smearing curves on the lattice spacing.

\section{Influence of the smearing strength on the continuum extrapolation}

\begin{figure}[ht]
\centering
\includegraphics[width=0.495\textwidth]{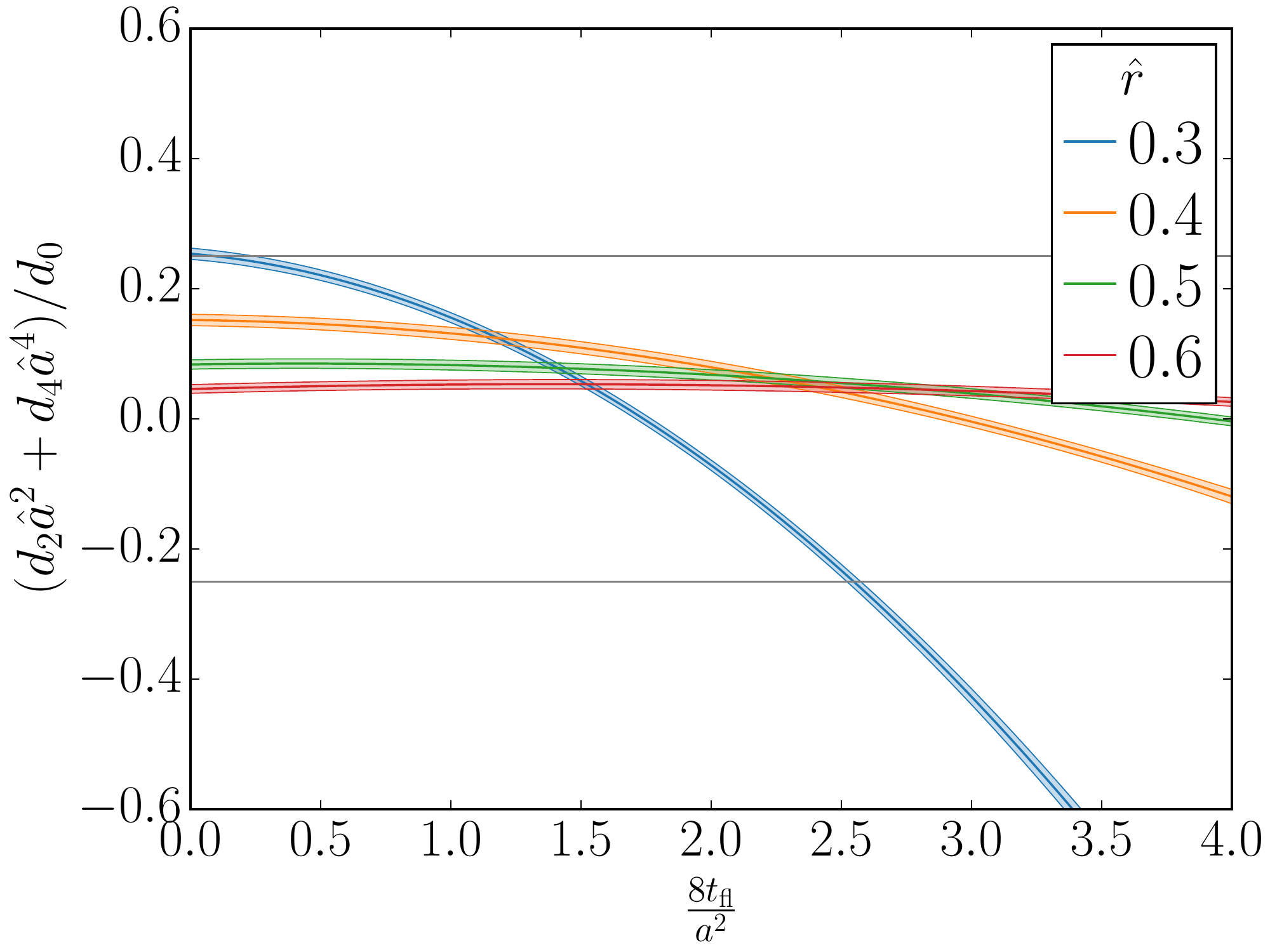}
\includegraphics[width=0.495\textwidth]{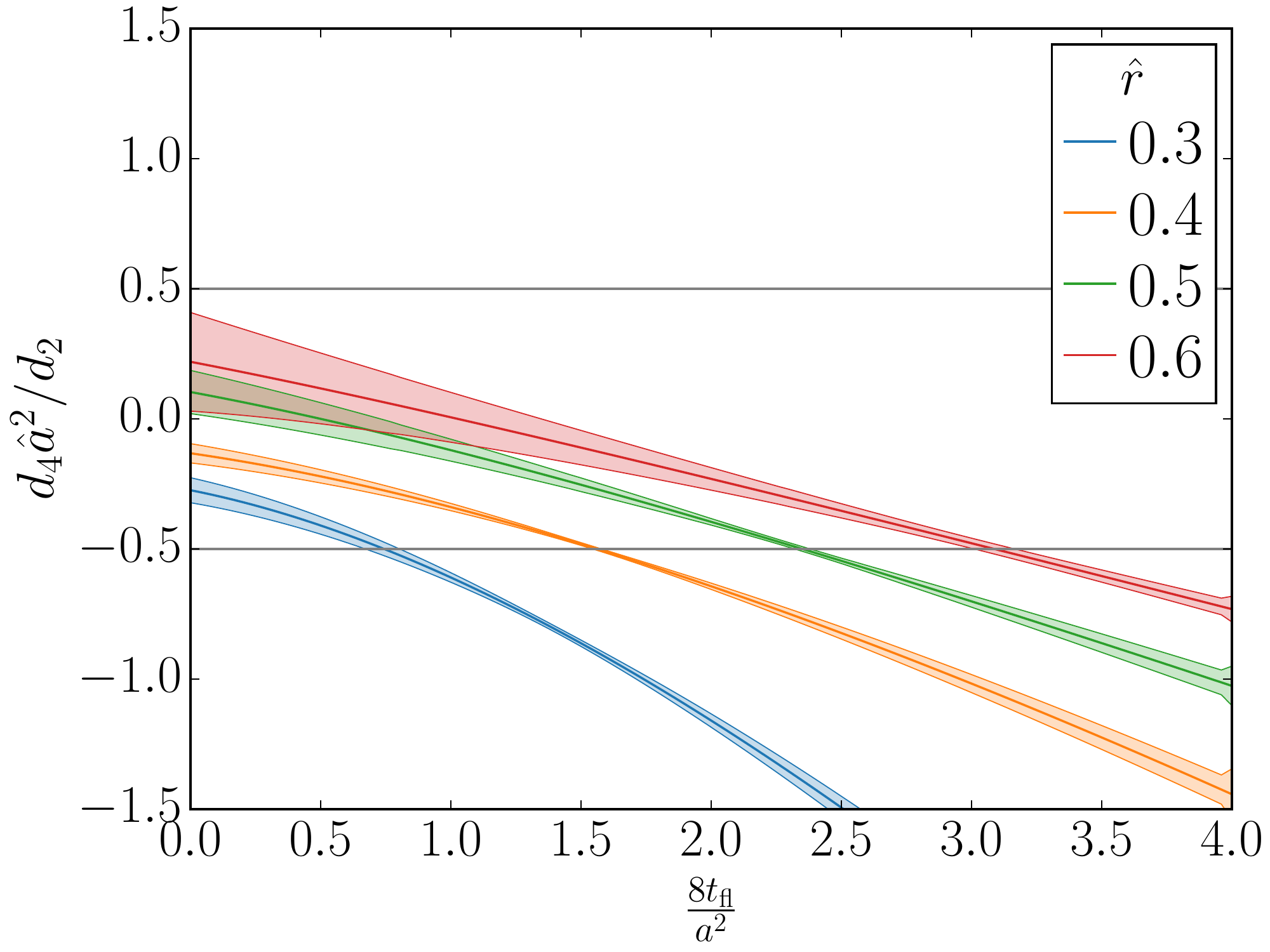}
\caption{Relative contributions to the Symanzik expansion of $\chihat\equiv\chi\cdot 8t_{0}$ evaluated at $a=0.062\,\fm$ (sft2) as a function of the smearing strength $\frac{8\tflow}{a^{2}}$ at various distances $\rhat\equiv\frac{r}{\sqrt{8t_{0}}}$.}
\label{fig:symexp}
\end{figure}
In order to investigate whether we are able to perform a controlled continuum extrapolation we consider relative contributions to the Symanzik expansion. We choose an intermediate lattice spacing (for some contemporary lattice computations this is a small one) of $a=0.062\,\fm$ . The relative total lattice artefacts $(d_{2} \ahat^2+d_{4}\ahat^4)/d_{0}$ and the $a^4$ contributions relative to the $a^2$ ones, $d_{4}\ahat^{2}/d_{2}$, are depicted in \cref{fig:symexp} as functions of the smearing strength $\frac{8\tflow}{a^{2}}$ for several distances $\rhat$. At smaller distances we find larger overall and smearing dependent lattice artefacts whereas at larger distances smaller overall lattice artefacts are present. In order to define a maximally allowed smearing strength we demand that the relative overall lattice artefacts are bounded by $|(d_{2}\ahat^2+d_{4}\ahat^4)/d_{0}|\leq 0.25$ and that the curvature is limited by $|d_{4}\ahat^2/d_{2}|\leq 0.5$, i.e. that lattice artefacts add up to less than $25\%$ and that the $a^4$-contribution is somewhat suppressed by the $a^2$ term. Note that with the signs of the coefficients found for the Creutz ratios, the second condition means that we are to the left side of the peak of the curves, i.e. in the region where lattice artefacts are monotonic. Clearly these conditions are not very stringent; one may want to insist on sharper bounds. From \cref{fig:symexp} we can deduce that our loose condition means $\frac{8\tflow}{a^{2}} \lessapprox 1$ for $\rhat=0.3$ while $\frac{8\tflow}{a^{2}} \approx 3$ is tolerated by the loose criteria at the upper end of the investigated distances, $\rhat=0.6$. At such distances the main restriction comes from the bound on the curvature.  

\begin{figure}[h]
\centering
\includegraphics[width=0.5\textwidth]{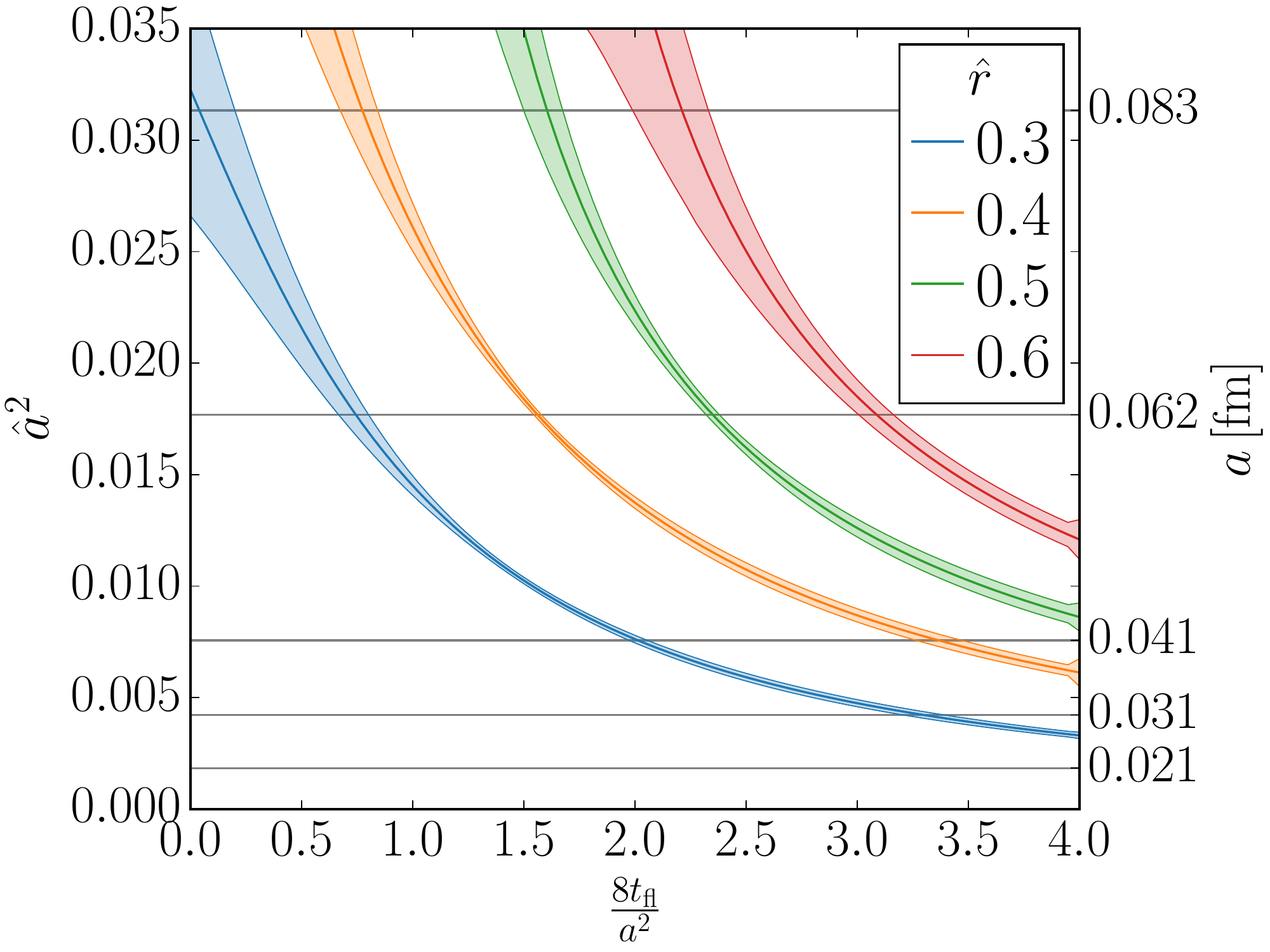}
\caption{Location of the maximum of $\chihat(\ahat)$ as a function of the smearing strength $\frac{8\tflow}{a^{2}}$ for several distances $\rhat\equiv\frac{r}{\sqrt{8t_{0}}}$.}
\label{fig:peaklocation}
\end{figure}

The location of the peak $\ahat_\mathrm{peak}^2$ is plotted as a function of the smearing radius for various distances $\rhat$ in \cref{fig:peaklocation}. Given a certain distance that one wants to cover adequately in the sense that cutoff-effects are monotonic, one has to be below the plotted curve. E.g. with $t_\mathrm{fl}/a^2=1$ and for distances $\rhat\geq 0.4$, lattice spacings up to $0.08$~fm lead to a monotonic dependence on $a$. In order to cover smaller distances one needs either smaller lattice spacings or less smearing. 

\section{Conclusion and Outlook}

We have studied the influence of gradient flow smearing on diagonal Creutz ratios $\chi(r,r)$ evaluated at various distances. We have found that the maximum tolerable smearing radius that still allows for a controlled continuum extrapolation depends on the distance. As expected, for short distance observables less smearing is tolerable. The main result of our numerical investigation is summarised in \cref{fig:peaklocation}. Each curve yields the upper boundary of the region where discretisation effects are monotonic, which is a minimum requirement for extrapolating data reliably to the continuum. Obviously, non-monotonic behaviour would not be an issue if we knew that a simple form as in~\cref{eq:fitsmearing} was correct. But even the asymptotic small $a$ behaviour is complicated with log-corrections~\cite{Husung:2017qjz} and higher orders in $a$, making a simple functional form essential for a controlled extrapolation.

Of course, we have only considered one specific observable in our study, but it is a measure of the force between quarks. This observable allows to probe different distances and we expect that physical observables with propagating quarks will be affected by how close it is to the continuum limit. We remind the reader that quark propagation with a smeared Dirac operator (smearing in the action) is related to smeared Wilson loops (smeared observables) through the hopping parameter expansion. Creutz ratios are just the finite version of Wilson loops. 

A further restriction of our study is in principle that we use gradient flow smearing with a continuous evolution in $t_{\flow}$ while in an application one will rather use stout smearing~\cite{Morningstar:2003gk}, which corresponds to a very coarse discretisation in $t_{\flow}$. We do not expect the behaviour in~\cref{fig:continuum} to change much but will check this explicitly in the near future.

\vspace{1em}
\begin{small}
Computations for this project have been performed on the HPC cluster PAX at DESY Zeuthen. The authors gratefully acknowledge the support of DV-Zeuthen.
\end{small}

\bibliographystyle{JHEP}
\bibliography{proceedings.bib}

\end{document}